\documentclass{jfm}

\usepackage[usenames, dvipsnames]{color}
\usepackage{graphicx}
\usepackage{epstopdf, epsfig}
\pdfoutput=1
\usepackage{amsmath}
\usepackage{amssymb}
\usepackage{cleveref}
\usepackage{multirow}

\usepackage{subcaption}

\begin{document}

\newtheorem{lemma}{Lemma}
\newtheorem{corollary}{Corollary}

\shorttitle{Impact force reduction} 
\shortauthor{R. Rabbi et. al.} 

\title{Impact force reduction by consecutive water entry of spheres}
\author{Rafsan Rabbi\aff{1},
Nathan B. Speirs\aff{2}, 
Akihito Kiyama\aff{1},
Jesse Belden\aff{3} \and
Tadd T. Truscott\aff{1}
\corresp{\email{taddtruscott@gmail.com}}
}

\affiliation
{
\aff{1}{Department of Mechanical and Aerospace Engineering, Utah State University, Logan, UT 84322}
\aff{2}{King Abdullah Univeristy of Science and Technology, Thuwal 23955, KSA}
\aff{3}{Naval Undersea Warfare Center, Newport, RI 02841}
}

\maketitle

\begin{abstract}
Free-falling objects impacting onto water pools experience a very high initial impact force, greatest at the moment when breaking through the free surface. Many have intuitively wondered whether throwing another object in front of an important object (like oneself) before impacting the water surface may reduce this high impact force. Here, we test this idea experimentally by allowing two spheres to consecutively enter the water and measuring the forces on the trailing sphere.  We find that the impact acceleration reduction on the trailing sphere depends on the dynamics of the cavity created by the first sphere and the relative timing of the second sphere impact.  These combined effects are captured by the non-dimensional `Matryoshka' number, which classifies the observed phenomena into four major regimes.  In three of these regimes, we find that the impact acceleration on the second sphere is reduced by up to 78\% relative to impact on a quiescent water surface. Surprisingly, in one of the regimes the force on the trailing sphere is dramatically increased by more than 400\% in the worst case observed.  We explain how the various stages of cavity evolution result in the observed alterations in impact force in this multi-body water entry problem. 
\end{abstract}


\section{Introduction} 
\label{Intro}
\begin{figure}
    \centering
    \includegraphics[width=\textwidth]{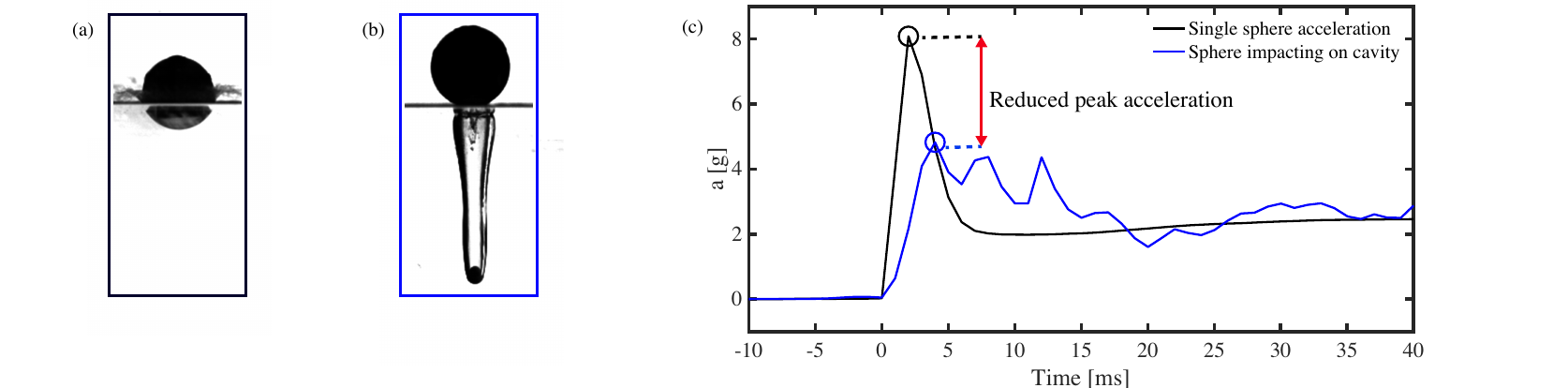}
    \caption{ Two different impact phenomena with their impact acceleration data plotted. (a) Free-surface water impact of a 50~mm vero plastic sphere ($U_o$~=~3.76 ms$^{-1}$). (b) The same sphere impacting at the same speed onto a cavity already created by a 10 mm sphere (3.38 ms$^{-1}$). (c) An accelerometer embedded in the vero plastic sphere reveals peaks of $\sim$8g for the case in (a) and $\sim$4.8g for the case in (b). The reduction in peak acceleration between the two indicates that the preformed cavity reduces the impact force on the trailing sphere.}
    \label{fig1}
\end{figure}

A prevailing myth is that water feels like concrete if one jumps onto it from a great enough height. Although this may seem like an oversimplification, the statement is somewhat truthful. The impact force felt at the time of penetrating a quiescent water surface can be very high \citep{Thompson1928, vonkarman1929, watanabe1933, Shiffman1945, May1975, Grady1979hydroballistics, Moghisi1981, korobkin1988}, much higher than the subsequent sustained underwater drag. One such example is shown in figure~\ref{fig1}(a,b), where a 50~mm sphere dropped from 0.72~m above the free surface results in an impulse with a peak impact acceleration of~$\sim$8$g$ whereas the underwater acceleration is close to a constant value of~$\sim$2g (\ref{fig1}(c)), indicating that at the moment of surface penetration the drag coefficient $C_d$ is four times the steady state underwater free-fall. This initially high impact force is primarily due to the large rate of change of momentum of the added fluid mass \citep{May1975,wang2019unsteady,wang2015analysis}, which is the highest during a submergence depth of 15-20$\%$ of the radius for spheres (figure~\ref{fig1}(a)) \citep{Shiffman1945,Moghisi1981}. Reducing this peak impact force is of significant interest because it presents structural failure risk to impinging bodies like aircraft landing on water, water landing spacecraft, underwater missiles, divers, base jumpers, etc.,\citep{kornhauser1964,May1975,seddon2006review, guillet2020hydrodynamics}. Previous studies have shown that impact forces can be reduced not only through object geometry \citep{mcgehee1959water, thompson1965dynamic,li1967study,may1970review,qi2016investigation,sharker2019,guzel2020reducing}, but also by modifying the near-surface region via, for example, aeration \citep{elhimer2017influence} or liquid jet-induced acceleration \citep{speirs2019jetball}.
An interesting extension to the idea of free surface modification is to launch a precursory object to agitate the free surface before entry. Such a concept has been proposed in popular culture (e.g., Mythbusters, Hollywood movies), yet has not received careful scientific investigation.

Here, we present the findings from a novel experiment investigating the consecutive water entry of two spheres, where the spheres are axially aligned and vertically separated (Figure~\ref{fig1}b). An accelerometer embedded in the upper sphere provides time-resolved measurements from which we deduce impact force. The lower sphere hits the water and creates a cavity through which the upper sphere falls, which can result in a reduced impact force on the upper sphere. Figure~\ref{fig1}c presents an example, where the upper sphere experiences a~$\sim$40$\%$ reduction in impact acceleration compared to the case where the same sphere impacts the quiescent free surface at the same velocity but without a cavity in front. We propose a modified version of the non-dimensional  parameter called the `Matryoshka' ($Mt$) number \citep{speirs2018water,hurd2015matryoshka} based on the cavity characteristics and the vertical spacing between the two spheres, which allows us to build an experimental regime diagram correlating different cavity conditions with the upper sphere impact force reduction and results in the observation of four distinct classes of consecutive two-sphere water entry behavior.

\section{Experimental Methods}

\begin{figure}
    \includegraphics[width=\textwidth]{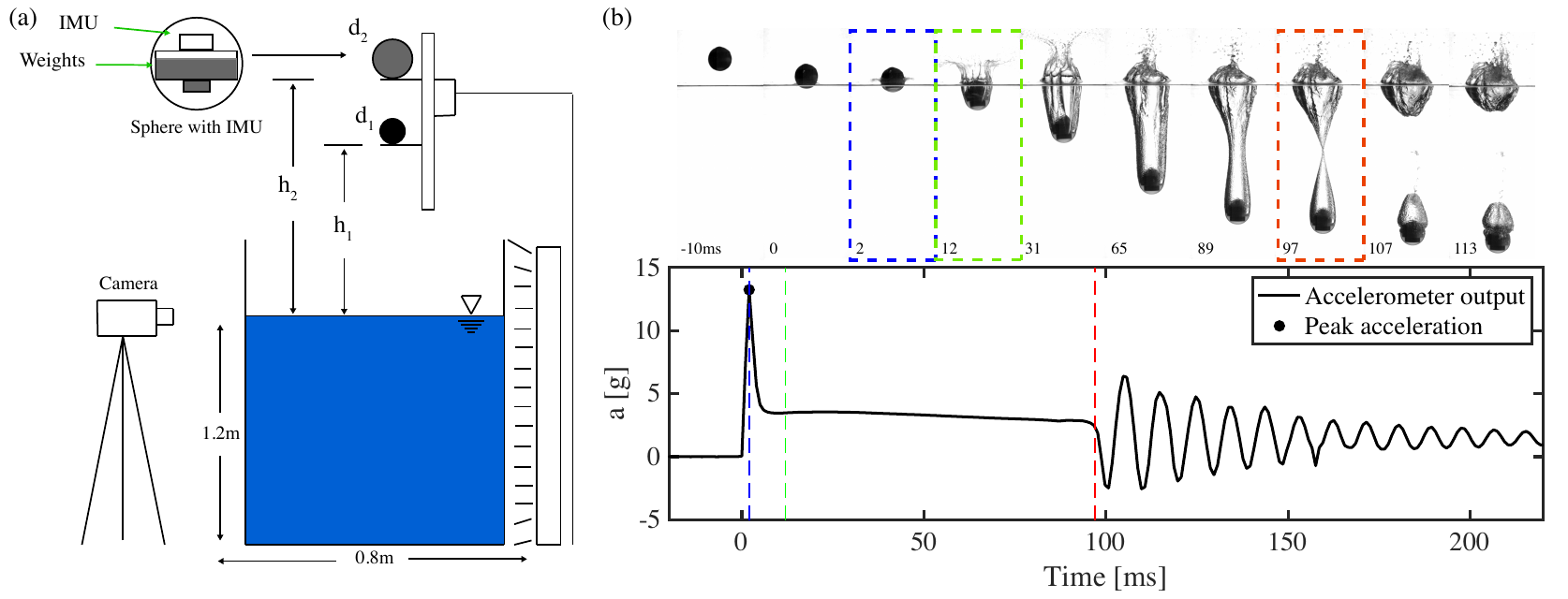}
\caption{A schematic diagram of the experimental setup for consecutive two-sphere water entry is shown in (a). The spacing $(h_2-h_1) = \Delta h$ is varied to attain different modes of two-sphere impact. (b) shows a quiescent drop where the upper sphere impacts the undisturbed water pool from height $h_2 = 0.72$~m and creates a cavity. The data taken by the accelerometer inside the sphere shows the acceleration at different stages of the sphere impact and entry, with the blue, green and red colored dashed boxes and lines indicating the time of peak acceleration, steady state underwater drag and starting point of cavity pinch-off.}
\label{fig2}
\end{figure}

Figure~\ref{fig2}(a) illustrates the experimental setup used for this study. 
Two spheres of diameter $d_2$ and $d_1$ were placed on two vertically separated axially aligned platforms held above a glass water tank. The platforms were kept parallel to the water surface with the help of a clamped string and pulley mechanism. When the platforms were let go, the two spheres would be in free-fall simultaneously and impact the water surface in tandem. The spheres were kept at heights of $h_2$ and $h_1$ from the water surface, the spacing ($\Delta h = (h_2-h_1)$) between the spheres varied from
0.07~m~-~1.24~m. The upper sphere was a 3d-printed vero plastic sphere with a fixed diameter of $d_2$~=~50 mm. Weights were inserted in the upper sphere to make it bottom heavy resulting in an upper sphere density of 2290~kgm$^{-3}$. Five different diameter steel spheres ($d_1$~=~10~mm-38~mm, density~7800 kgm$^{-3}$) were used as the lower sphere, sprayed with Cytonix WX-2100 coating to make them hydrophobic, resulting in a surface contact angle of 117$^o$ and the increased the roughness of the spheres to $R_z =$~50.2 $\pm$ 21.4 $\mu$m (95\% confidence). The vero plastic upper sphere has a hydrophilic surface of wetting angle $\theta$ = 80 $\pm$  8$^o$ and surface roughness $R_{z} = 7.2 \pm 1.2 \mu$m (95\% confidence). 

The upper sphere housed an Inertial measurement unit (IMU) built in-house, with two three-axis accelerometers, one gyroscope and one magnetometer embedded. The two accelerometers on-board were one low-range and one high-range. The low range accelerometer has a measurement range of $\pm$16g, it is a MPU-9250 motion tracking device manufactured by Invensen Inc. The high range accelerometer is a chip called H3LIS331DL produced by ST, and was set to a maximum range of $\pm$100g. Both would register data for any drop event, whenever possible the data from the low accelerometer is reported, because the high accelerometer is more prone to noise. Data from both are comparable for cases where acceleration values were within $\pm$16g. The accelerometer sampling rate is limited to 1000~Hz. The root sum square of the acceleration values from three axis is calculated and reported as the total acceleration. 

Figure~\ref{fig2}(b) shows a typical quiescent upper sphere impact event, with the acceleration output from the IMU shown in figure~\ref{fig2}(c). The sphere impacts the free-surface at 0ms, and a sudden increment of acceleration is registered. The peak acceleration is reached soon after, shown by the black dot and the time marked by blue dashed line. This impact pulse lasts fleetingly until $\sim$8~ms, after which the sphere travel downwards with an air cavity in its' wake until 97~ms, when the .

\section{Results and discussion}
\subsection{Scaling analysis and formulation of `Matryoshka' number}
When an object impacts a water pool it displaces some of the water with air and accelerates fluid downwards as the object falls through the pool, leaving an air filled cavity in its wake \citep{Truscott2014}.  Creating a cavity in front of an impacting object can be conducive to reducing its impact acceleration, as evident from figure~\ref{fig1}c. The state of the cavity over time indicates the local liquid flow-field surrounding the cavity \citep{truscott2012unsteady,mansoor2014water}, which may help explain the change in impact acceleration for any trailing object.  Thus, understanding cavity creation and evolution is paramount for determining why and how an air cavity may reduce impact force. 

\begin{figure}
    \centering
    \includegraphics[width=\textwidth]{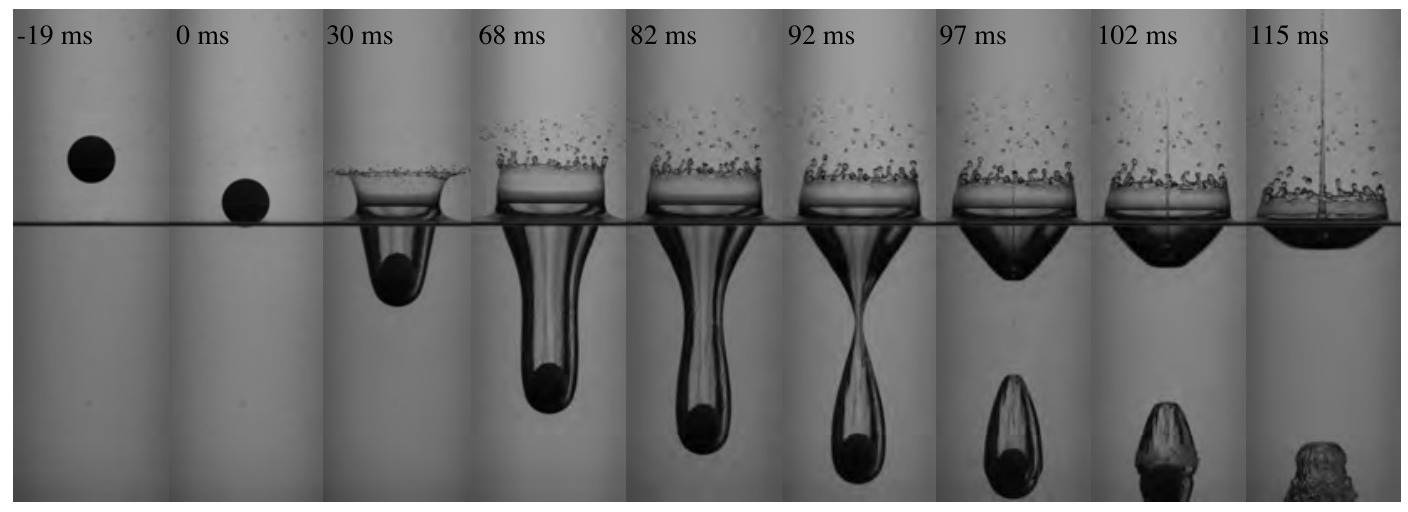}
    \caption{ A typical lower sphere impact on quiescent water surface without a trailing upper sphere. A 38 mm hydrophobic steel sphere impacts a quiescent water pool ($U_o\sim$~3.2 ms$^{-1}$) creating a subsurface air cavity (0~-~68~ms). The cavity elongates in time with the downwards moving sphere (30~-~92~ms) and finally collapses on itself in a deep seal (92 ms) pinch-off. After the deep seal, the cavity is divided in two parts, a cavity bubble attached with the sphere moving downwards (97~ms), and the upper bowl shaped distortion of free surface which eventually creates a Worthington jet (97~-~115 ms).}
    \label{fig3}
\end{figure}

Objects with rough and hydrophobic surfaces almost always create cavities even at very low impact velocities \citep{Duez2007,Zhao2014,speirs_mansoor_belden_truscott_2019}. Figure~\ref{fig3} shows such a case where a 38~mm hydrophobic sphere 
creates an axisymmetric cavity at an impact velocity of $\sim$3.2~ms$^{-1}$. The cavity elongates with the downward moving sphere, until the point when hydrostatic pressure forces the cavity to seal near the cavity mid-point at 92~ms. This sealing event is popularly referred to as `deep-seal' pinch-off. After pinch-off the cavity divides into two parts, a pulsating air bubble attached to the downwards moving sphere and the upper bowl shaped distortion in the free surface retreating upwards creating a high-speed axisymmetric 'Worthington' jet \citep{Worthington1897, gekle2010}, as seen in figure~\ref{fig2}a from 92~-~115~ms. Different impact velocities and sphere sizes result in different cavity behaviors which can be classified by cavity seal type
\citep{Aristoff2009,speirs_mansoor_belden_truscott_2019}.  For example, the cavity shown in figure~\ref{fig3}a is referred to as deep seal cavity because of the characteristic mid depth deep seal pinch-off. 

\begin{figure}
    \centering
    \includegraphics[width=\textwidth]{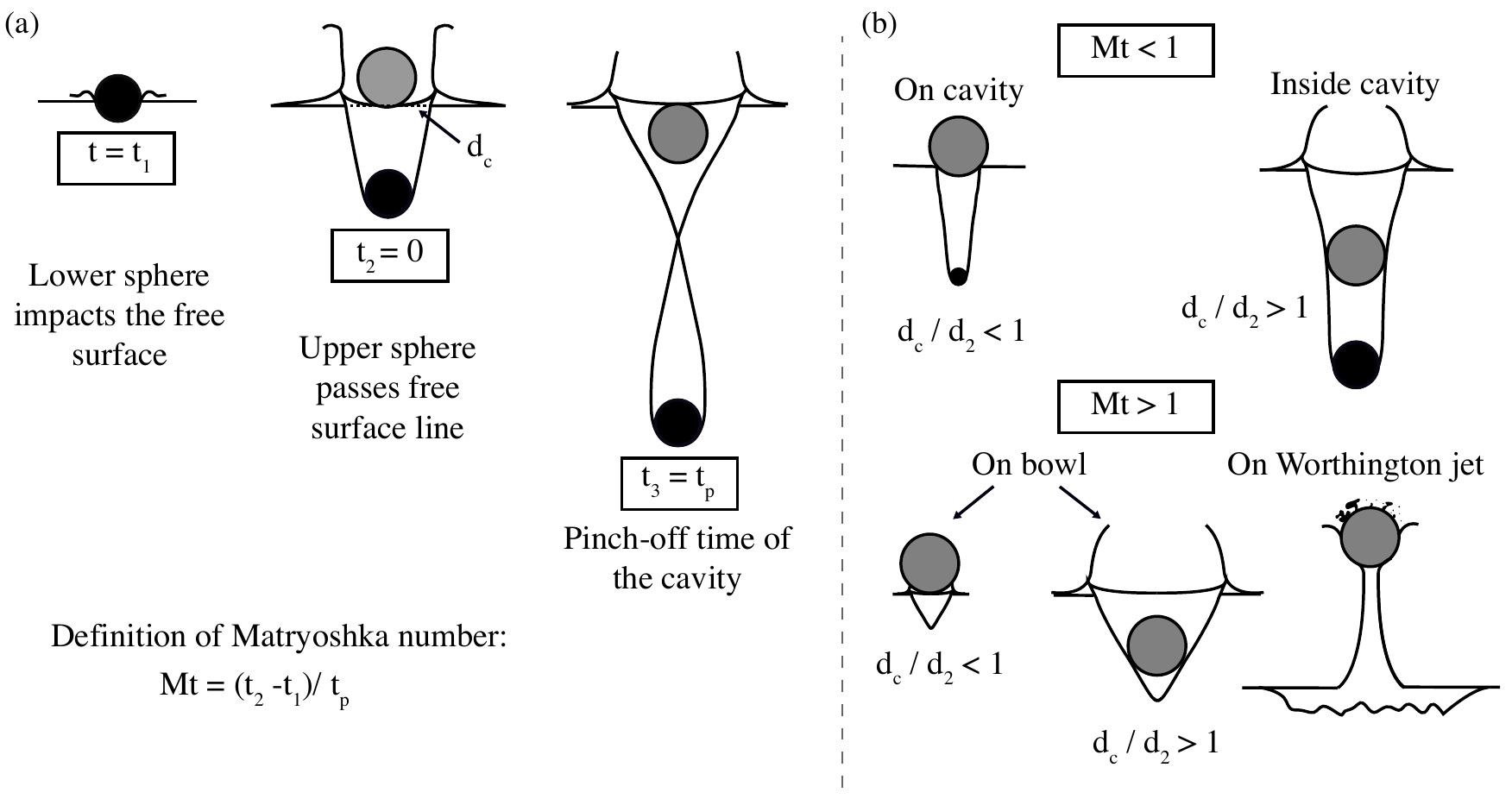}
    \caption{ The formulation of non-dimensional `Matryoshka' number and and predicting different modes of two-sphere water entry using cavity pinch-off time information. (a) When two spheres vertically separated by a distance $\Delta h$ are let go from their rest position, the variation in spacing will result in the upper sphere impacting various stages of the cavity created by the lower sphere. Taking the characteristic deep seal pinch-off time as $t_p$, a non-dimensional time parameter called `Matryoshka' number can be defined. (b) $Mt = 1$ denotes boundary between different modes of consecutive two-sphere impact. For $Mt<1$, the upper sphere can interact with a still growing cavity, yielding cavity cases: `on cavity' and 'inside cavity'. Beyond $Mt>1$, the upper sphere should interact with the collapsed cavity, resulting in two different `on bowl' modes based on the cavity opening diameter $d_c$ or with the Worthington jet ('on jet' cases).}
    \label{fig4}
\end{figure}

\begin{figure}
    \centering
    \includegraphics[width=\textwidth]{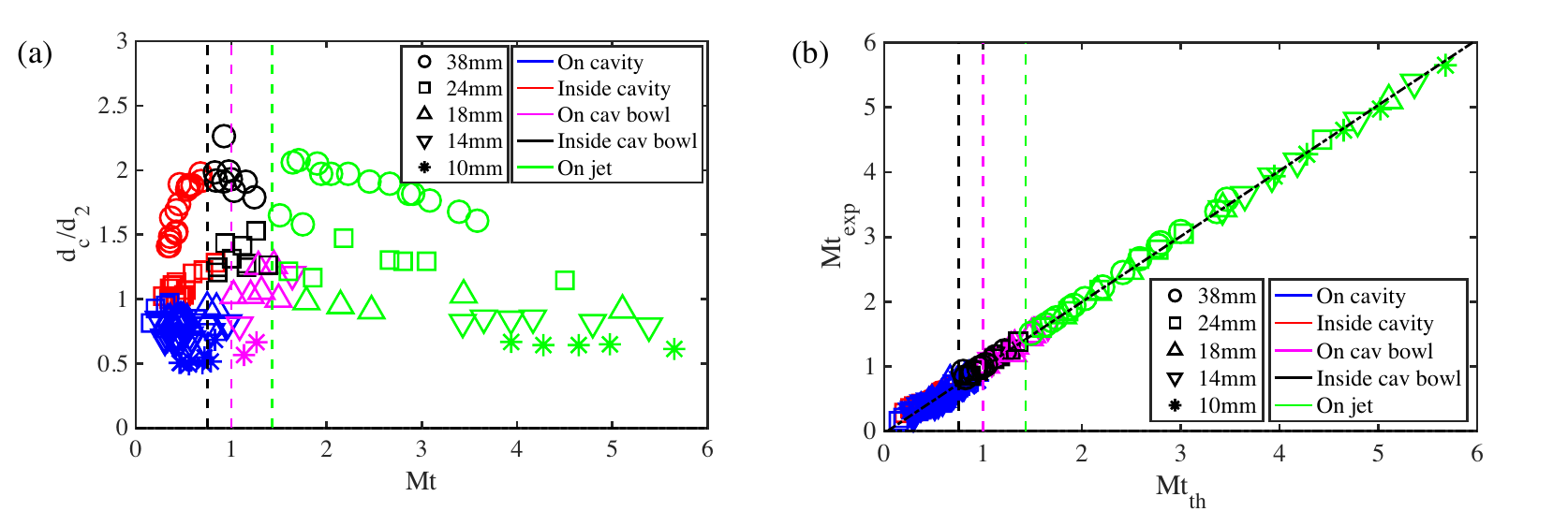}
    \caption{ (a) The experimental map of Mt vs ratio of cavity to the upper sphere diameter. The different colored markers points out to different two-sphere entry modes, with red, blue, magenta and green denoting inside cavity, on cavity, on bowl and on jet cases respectively. (b) shows the approximately one to one relation between experimentally and theoretically calculated $Mt$. The dash-dotted black line illustrates the experimental fit line ($Mt_{exp}=1.013Mt_{th}-0.02956$). The dashed black, magenta and green vertical lines at 0.75, 1 and 1.43 respectively show the transitions between cavity cases (on cavity and inside cavity) and non-cavity cases (on bowl and on jet) in (a, b).
    }
    \label{fig5}
\end{figure}

Increasing the impact velocity for the same sphere sizes results in surface seal \citep{mansoor2014water,Aristoff2009, speirs_mansoor_belden_truscott_2019}, which is distinguished by the splash-crown sealing above the free-surface, and the resulting detachment and pull-away of the cavity below the free surface. 
In the context of consecutive two-sphere water entry, one might expect the pinch-off (or seal) event from the cavity of the first sphere to affect the dynamics of the trailing sphere.  For two axially aligned, vertically separated spheres (upper sphere diameter $d_2$, lower sphere diameter $d_1$) as shown in figure~\ref{fig2}(a), varying the spacing $\Delta h = (h_2-h_1)$ between the two spheres will result in the upper sphere interacting with the cavity either before or after pinch-off, which we anticipate will lead to different sphere-cavity interaction modes. Taking the pinch-off time as a characteristic time scale, we propose to characterize consecutive two-sphere water entry with a modification of the non-dimensional parameter known as the `Matryoshka' number $Mt$.  This term has been used in prior research to describe successive cavity formation from multi-droplet impacts using droplet frequency and cavity formation time as the fundamental timescales \citep{hurd2015matryoshka, speirs2018water}.  In a physical sense, $Mt$ can be considered a ratio of the time to completion of a single event to the consecutive initiation of the same event by the second sphere.  Here, we define
\begin{equation}
    Mt = \frac{\Delta t}{t_p}.
    \label{eq1}
\end{equation}
where $\Delta t = |t_2 - t_1|$ is the time difference between the two spheres passing the free-surface (figure~\ref{fig4}a), and $t_p$ is the pinch-off time of the first cavity. Thus, $Mt$ parameterizes the state of the cavity formed by the first sphere at the time when the second sphere interacts with it. $Mt<1$ indicates the first cavity has not gone through pinch-off, which results in cases where the upper sphere interacts with an elongating cavity. For cavity opening diameter (itself a function of time and the lower sphere diameter \citep{Duclaux2007, Aristoff2009}) $d_c < d_2$, the upper sphere falls on the cavity when impacting the water pool, and we name these `on cavity' cases (figure~\ref{fig4}b). When $d_c > d_2$, then the upper sphere falls through the cavity opening, which we call the `inside cavity' case, with $d_c/d_2 = 1$ working as the transition between the two cavity cases predicted for $Mt = 1$ (figure~\ref{fig5}(a)). 
For $Mt>1$, the trailing sphere interacts with the upper detached portion of the cavity, either falling through ($d_c > d_2$) or falling on ($d_c < d_2$) the bowl shaped, retreating free surface; or falling through a Worthington jet resulting from cavity pinch-off at higher $Mt$ (figure~\ref{fig4}d). We call these cases  `on bowl' and `on jet',  respectively.

The value of $Mt$ can be predicted \textit{a priori} provided a prediction of the pinch-off time, since $\Delta t$ can be predicted using the spacing between the two spheres. For low Froude number $(Fr = U_o^{2}/gd_{s}$, $ U_o$ is the impact velocity, $d_s$ is the sphere diameter and $g$ is the gravitational constant) deep seal pinch-off time can be written as $t_p = \beta \sqrt{d_s/2g}$, \citep{glasheen1996vertical,Duclaux2007,truscott2009water}, where $\beta$ is an experimental constant with different values ranging from 1.72 to 2.285 proposed in the literature \citep{bergmann2009controlled,Duclaux2007,marston2012cavity}. In this study, experimentally calculated $\beta$ = 2.03$\pm$0.0974 (95\% confidence, found experimentally) is used (figure~\textcolor{red}{S?}).
For higher $Fr$ where surface seals are expected to happen, this equation overestimates the pinch-off time marginally (figure~~\textcolor{red}{S?}), but the lack of a good consensus in literature about the surface seal time and the scale of the experiments carried out in this paper ($Fr<400$) makes this equation a good approximation. To predict $\Delta t$, we calculate the spacing between the two spheres. The lower sphere impacts the quiescent water surface first at time $t=t_1$ (figure~\ref{fig4}a), travelling a distance $h_1$ (figure~\ref{fig2}a). The upper sphere also travels the same distance $h_1$ in that time and is at height $\Delta h = (h_2-h_1)$ from the free surface. At that moment, the upper sphere is travelling with a velocity of $u_1$, same velocity with which the lower sphere impacts the free surface. If the upper sphere takes time ($\Delta t = (t_2-t_1)$) to pass the free surface line from height $\Delta h$, then $\Delta h = u_1 \Delta t+0.5g\Delta t^2$. Solving this for $\Delta t$ yields $\Delta t = (-u_1+\sqrt{u_1^{2}+2g\Delta h})/2g$. substituting this in~\ref{eq1} and using $u_1=\sqrt{2gh_1}$,
\begin{equation}
    Mt_{th} = \frac{2(-\sqrt{h_1}+\sqrt{h_1+\Delta h}}{\beta \sqrt{d_1}})
    \label{Mteq2}
\end{equation}
 Figure~\ref{fig5}(b) plots the experimentally measured $Mt_{exp}$ against their theoretically calculated counterpart. The black dotted experimental fit line shows good agreement between experimentally and theoretically calculated $Mt$.
 A natural question might arise here over the use of $Mt$ as the scaling for consecutive two-sphere impact. Since $Mt$ depends on the time difference between consecutive impact, it would be prudent to ask ourselves whether only using the spacing $\Delta h = (h_2-h_1)$ between the spheres should be an easier scaling to follow because the time difference to impact depends on this distance. Although tempting, it ultimately doesn't hold up in providing all the information necessary for predicting two-sphere impact and resultant force reduction. A very important piece of this puzzle is the sphere size. Since the pinch-off time and the cavity opening diameter are a function of sphere diameter, defining $Mt$ with the pinch-off time in the formulation does better as a scaling parameter(figure~\textcolor{red}{S?}).

\subsection{Consecutive sphere entry: different modes}

\begin{figure}
    \centering
    \includegraphics[width=\textwidth]{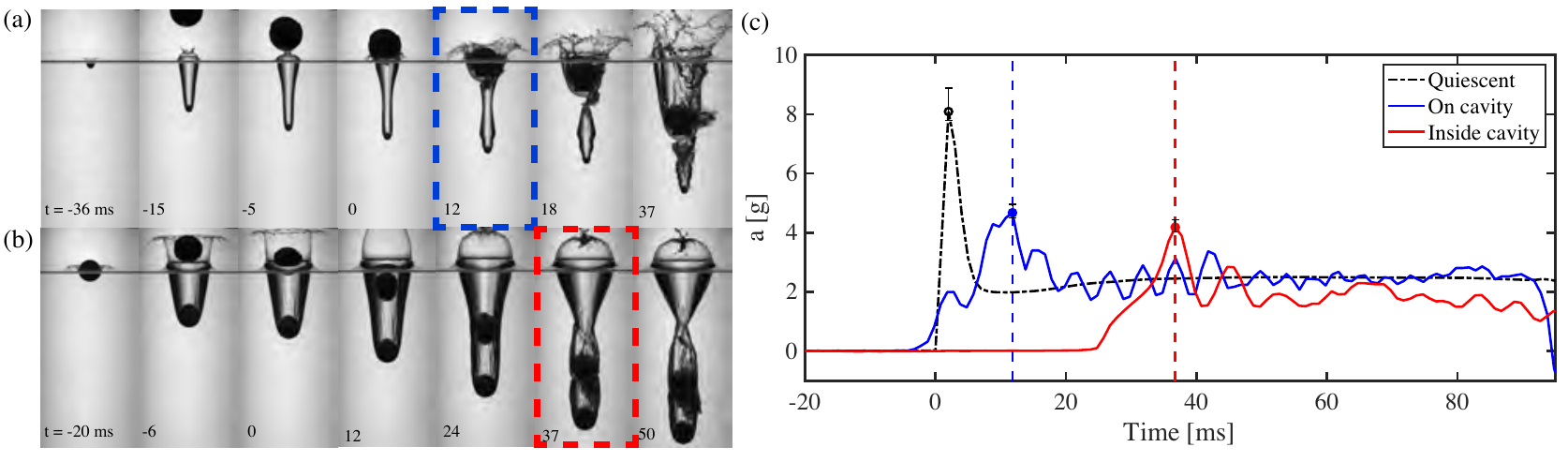}
    \caption{ Time series for two example cavity cases in consecutive sphere entry. A hydrophobic sphere ($d_1$ = 10~mm (a) and 38~mm (b)) impacts the quiescent free surface and creates a cavity, followed by a 50~mm sphere with embedded IMU in free fall from a height of $h_2$~=~0.72~m, where $\Delta h$~= 0.14 and 0.11 respectively for (a,b). The time the upper sphere passes the free surface line is considered as t = 0~ms for each of the cases, yielding $Mt\sim0.82 \& \sim0.35$ for the on cavity (a) and inside cavity (b) cases. The blue and red dashed lines in (a,b \& c) denote the time of peak acceleration felt by the upper sphere, significantly less than the quiescent value of~8g.}
    \label{fig6}
\end{figure}

\begin{figure}
    \centering
    \includegraphics[width=\textwidth]{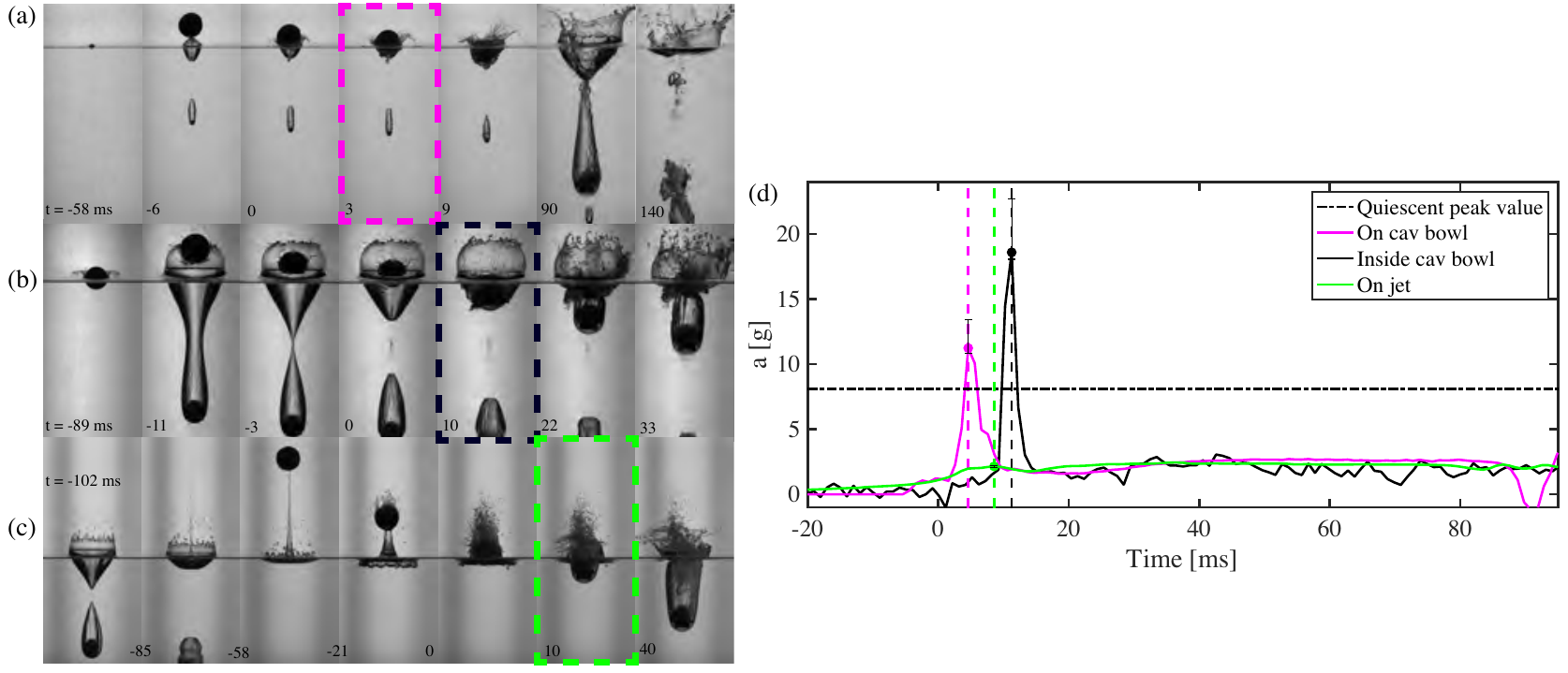}
    \caption{ 
    A hydrophobic sphere ($d_1$ = 10~mm (a) and 38~mm (b \& c)) impacts the quiescent free surface and creates a cavity, followed by the 50~mm upper sphere in free fall from a height of $h_2$~=~0.72~m, where the cases are: (a) inside cavity bowl is $\Delta h$~= 0.22 \& $Mt\sim1.26$, (b) on cavity bowl is  $\Delta h$~= 0.3 \& $Mt\sim1.05$ and (c) on jet is $\Delta h$~= 0.54 \& $Mt\sim2.20$. 
    (d) acceleration of each of the cases in a-c with peak acceleration marked in both with a dashed line. 
    }
    \label{fig7}
\end{figure}

As explained in the previous section, $Mt<1$ results in cavity cases (on cavity and inside cavity) depending on the ratio $d_{c}/d_{2}$ (figure~\ref{fig4}(b),~\ref{fig5}(a)), and $Mt>1$ indicates the non-cavity cases (on bowl and on jet). Figure~\ref{fig6} \& figure~\ref{fig7} presents time-series murals of all two-sphere modes including two different on bowl cases (figure~\ref{fig7}(a,b), with their dynamic acceleration response plotted with the murals. The accelerations at impact are reduced for both on cavity and inside cavity cases compared to the quiescent case peak. In the on bowl cases, the initial impact pulse has a \textit{higher} peak than the quiescent case, indicating a higher impact force experienced during free-surface entry. For the on jet cases,  the peak acceleration value is significantly smaller than the quiescent peak. The time of the peak accelerations for the two-sphere cases happen later than the quiescent case (see figure~\ref{fig6}(a-c),~\ref{fig7}(a-d))  
, since the sphere interacts with modified free surface conditions.

\begin{figure}
\centering
\includegraphics[width=0.7\textwidth]{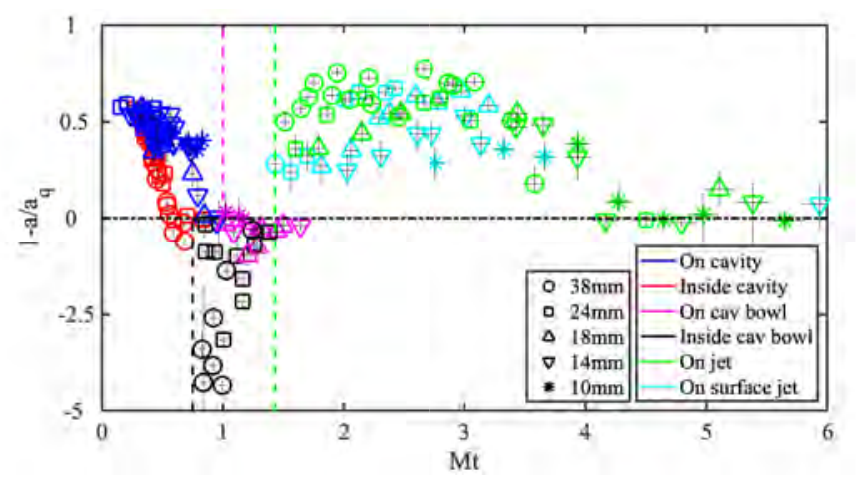}
    \caption{ Percentage reduction in the initial impact force for the upper sphere versus {\it Mt} for different two-sphere water entry modes. 
    The transition from inside cavity to on bowl occurs at $Mt\sim$0.75 (black dashed line) and on cavity to on bowl mode at $Mt\sim$1 (magenta dashed line).
    The on jet cases are split into two cases: a clean single upper sphere impact onto a Worthington jet (on jet mode) as shown in Figure~\ref{fig7}(c); and a variation where the splash dome-over at the free surface suppresses the Worthington jet creating a small jet-like water column at the surface we term as `on surface jet'.  The horizontal black dashed-dot line at $1-a/a_q = 0$ in the y-axis indicates $a = a_q$, while the vertical green dashed line represents the experimentally found transition from both on bowl modes to both on jet modes.  The error bars are marked with 95\% confidence with more information available in \textcolor{red}{SI section B}. 
    }
\label{fig8}
\end{figure}

Figure~\ref{fig8} shows a regime diagram where reduction in acceleration for varying $Mt$ is plotted for the range of experimental conditions. Reduction in acceleration is computed as $1-a/a_q$, where $a$ is the measured peak acceleration of the trailing sphere in a two-sphere water entry, and $a_q$ is the peak acceleration of the same sphere impacting quiescent water from the same drop height ($h_2$). Similar to the standalone cases presented in figure~\ref{fig6},~\ref{fig7}, on cavity ($0<Mt<1$) and inside cavity cases ($0<Mt<0.75$)  experience notable reduction in impact acceleration, with a downward linear trend in reduction values present for both. Alternatively, the on bowl cases experience an increase in impact acceleration, evident from the negative reduction in acceleration in figure~\ref{fig8} up to $Mt\sim1.43$. The on jet cases experience dramatic reduction in impact acceleration with a downward trend from $1.43<Mt<6$.

This reduction-gain-reduction trend in peak impact acceleration seen in figure~\ref{fig4} can be explained considering a force balance in which a drag term quadratic in velocity is dominant.  For a sphere of diameter $d_s$ entering quiescent water, this force balance yields,
\begin{equation}
a_{q} =\frac{\frac{1}{2} C_{d} A_{s} \rho U^{2}_{o}}{\rho _{s} V_{s}}
\label{eq2}
\end{equation}
Where $C_d$ is the drag coefficient, $A_s$ is the effective frontal area for the sphere, $\rho$ and $\rho_{s}$ are the water and sphere density respectively, $U_o$ is the impact velocity and $V_s$ is the volume of the sphere. For consecutive two sphere impact, the lower sphere creates a cavity in front of the upper sphere, changing the effective frontal area to $A_{e}$ and accelerating the fluid in the pool with the elongating cavity resulting in a relative impact velocity of the upper sphere $U_{rel}$. These effects culminate in the modified version of~ Eq.\ref{eq2},
\begin{equation}
a_{q} =\frac{\frac{1}{2} C_{d} A_{e} \rho U^{2}_{rel}}{\rho _{s} V_{s}}
\label{eq3}
\end{equation}

The impact acceleration increases with increasing $Mt$ for on cavity and inside cavity cases (figure \ref{fig8}).
 This is expected taking into account the evolution characteristics of the flow field surrounding subsurface air cavities with time and its effect on Eq.\ref{eq3}. We must first look at the first sphere entry. Early in the impact, the cavity is quickly expanding radially and decreases in expansion rate as the cavity approaches  pinch-off and begins to radially collapse (Figure~\ref{fig3}, \textcolor{red}{S8}(a)). Thus, for a smaller $Mt$ in the on cavity regime, the sphere enters the cavity when it is radially expanding and the velocity field near the cavity surface is directed outward with relatively large magnitude \citep{truscott2012unsteady, mansoor2014water}. This makes the relative impact velocity $U_{rel}$ of the upper sphere for on cavity cases much lower, resulting in higher reduction of acceleration (i.e., force) as predicted by equation~\ref{eq3}. As $Mt$ increases in the on cavity regime, the second sphere enters the cavity when the velocity field close to the free surface is moving inward, thus increasing the relative velocity of impact, which in turn results in a lower reduction of force. Overall, the on cavity cases follow a downward linear trend. Furthermore, the upper sphere falls onto an air gap formed on the free surface by the cavity for both on cavity and inside cavity modes, which also contributes to the reduction of impact acceleration. For the inside cavity cases, the relative velocity argument also holds true until the cavity starts collapsing on itself. After falling through the cavity opening, the upper sphere impacts the inside cavity wall at depths where the cone-shaped cavity has narrowed to the diameter of the sphere. As $Mt$ increases, the depth at which the upper sphere impacts the walls increases, and the impact happens closer in time to pinch-off. At these deeper locations, hydrostatic pressure forces the cavity back inward resulting in a greater relative velocity and increased amount of fluid the upper sphere is in contact with \cite{Duclaux2007,Aristoff2009,speirs_mansoor_belden_truscott_2019}. Consequently ,the upper sphere experiences progressively higher acceleration at impact for greater $Mt$ numbers.
 It should also be noted that the slope of force reduction is much steeper for inside cavity cases compared to on cavity cases.
 
 The cases where the second sphere enters the wake of the first after pinch-off should happen beyond $Mt>1$ as predicted in figure~\ref{fig4}(b). Note that in theory both on cavity and inside cavity cases should extend to $Mt\sim1$, but in reality only the on cavity cases transition to the next mode at the predicted $Mt\sim1$ juncture. In observation, $Mt=0.75$ is the transition criterion for the inside cavity cases.The transition occurs earlier for the inside cavity cases because of the condition $d_c>d_2$, which means for inside cavity cases the upper sphere falls through the cavity when passing the free surface. Thus, there is a  possibility for such cases to exist where the sphere passes the free surface when the cavity hasn't undergone deep-seal pinch-off, but before the sphere impacts the narrowing cavity wall pinch-off occurs. The sphere crashes through the pinch-off singularity or the upward moving bowl in these unique cases and we label them on bowl cases. We can analytically predict the $Mt$ where this transition happens by calculating the time it takes for the sphere to travel from free surface to the pinch-off location ($t_{HP}$).
 If the sphere passes the free surface line at time $t_2=0$ after the lower sphere impacts the free surface (time $t_1$), then we can write the pinch-off time $t_p$ as, 
\begin{equation}
    t_p = (t_2-t_1)+t_{HP},
    \label{eq4}
\end{equation}
Dividing both sides of Eq.~\ref{eq4} by $t_p$  and rearranging we get,
\begin{equation}
    Mt^{*} = 1-\frac{t_{HP}}{t_P}
    \label{eq5}
\end{equation}
The time $t_{HP}$ can be predicted using the relation $t_{HP} = {v-u}/{g}$, where $v$ is the velocity with which the upper sphere impacts the pinch-off point and $u$ is the initial velocity with which the upper sphere passes the free surface ($u = \sqrt{2gh_2}$). The value $v$ can be predicted by the relation $v = \sqrt{u^2+2gH_p}$, where $H_p$ is the pinch-off depth. $H_p$ can be predicted using the relation $H_p = 0.4d_{1}Fr^{1/2}$ \citep{Duclaux2007}, where $d_1$ is the lower sphere diameter. Then, Eq.~\ref{eq5} can be rewritten as,
\begin{equation}
    Mt^{*} = 1-\frac{\sqrt{u^2+2gH_p}-u}{gt_p}\notag\\
    = 1-\frac{2(\sqrt{h_2+0.4d_{1}Fr^{1/2}}-\sqrt{h_2})}{\beta\sqrt{d_1}}\tag{3.8}
    \label{eq6}
\end{equation}
 Where $t_p$ can be written as $t_p = \beta \sqrt{d_1/2g}$. Using the different impact conditions for the inside cavity experiments, we can use Eq.~\ref{eq6} to predict a transition $Mt^{*}$ where the upper sphere falls right on the pinch-off singularity. Using 33 different two-sphere cases, this $Mt^{*}$ is determined to be 0.75$\pm$0.015 (95\% confidence). Plotting this line as the separation criterion for inside cavity and on bowl cases in figure~\ref{fig4} illustrates that the experimental data agrees with this theoretical $Mt = 0.75$ separation line quite well.

Depending on the cavity opening diameter $d_c$, two different on bowl modes are possible: on cavity bowl ($d_c<d_2$) and inside cavity bowl ($d_c>d_2$). They transition from on cavity and inside cavity modes at $Mt\sim1$ and $\sim0.75$ respectively as discussed earlier. In both, the cavity bowl introduces an upward velocity field in front of the impacting upper sphere, resulting in a larger value of $U_{rel}$ and hence a larger force of impact (\textcolor{red}{figure S9}). In addition, this bowl is in the shape of a deformed free surface, which in many cases has a greater curvature 
causing the effective frontal area of impact $A_{e}$ to increase in certain on bowl cases (figure~\ref{fig7}(a), 0~-~9~ms,~\ref{fig7}(b), 10~-~20 ms). Thus, the upper sphere impacts and penetrates a water surface with high curvature and upward velocity, the coupled effects of both these phenomena leads to a dramatically higher peak impact acceleration than a normal quiescent case (figure~\ref{fig3}f). Figure~\ref{fig8} shows that the on bowl cases almost always have higher impact acceleration, in some instances impact forces almost quadruple that of the quiescent case value ($\approx 427\% $ at $Mt\sim 0.96$), making this $0.75<Mt<1.43$ range of Matryoshka values a range to avoid if one wishes to achieve any sort of reduction in impact force. The inside cavity bowl cases show much higher accelerations than their on cavity bowl counterparts, which can be attributed to the higher bowl retraction speed observed in cavities created by larger diameter spheres ($d_c>d_2$).  

The bowl eventually forms an upward moving axial jet called a Worthington jet \cite{Worthington1897, Worthington1908,gekle2010} coming out of the base of the distorted bowl-shaped free surface following the collapse of the cavity after pinch-off ($Mt\geq1.43)$.
Here, the on jet cases are characterized by the upper sphere dramatically passing through the Worthington jet into the water pool (figure~\ref{fig7}(c)). We can predict the onset of the on jet cases by estimating the time it takes for a Worthington jet to fully form after pinch-off as $t_j$ and modify the Matryoshka number as $Mt^{**} = 1+t_j/t_p$ (see section SI \textcolor{red}{F}). The averaged $Mt^{**}$ onset from 20 experiments turns out to be 1.43$\pm$0.06 (95\% confidence). 

The acceleration reduction for these cases beyond $Mt\geq1.43$ are also dramatic, with the highest reduction of up to $\sim$78\% observed experimentally (Mt~=~2.67, figure~\ref{fig8}). The time steps and the acceleration plot presented in figure~\ref{fig7}(c) provide an explanation of how the reduction occurs. The jet starts wetting the trailing upper sphere long before it has reached the free surface (80 ms before free surface impact, not shown in figure~\ref{fig7}(d) on jet case, see supplemental figure~\textcolor{red}{S12}). This drawn out collision with the narrow axial jet results in a reduction in momentum of the upper sphere over a longer period of time, and also results in partial wetting of the upper sphere by the jet. At the time of impact, the water-enveloped upper sphere doesn't abruptly go through an air-water interface like the quiescent case does. Instead, it enters the water partially wetted with reduced momentum. 
With the peak impact reduced, the sphere experiences accelerations similar to a free-falling sphere through water without any apparent impact event in the acceleration (figure~\ref{fig7}(c), on jet case, 0 ms). 

There is an overall downward trend in force reduction for $Mt>1.43$ in the on jet regime in figure~\ref{fig8}. A greater $Mt$ indicates a larger time difference between the jet creation and the upper sphere passing the free surface. As time between the two sphere free surface impact is increased, the Worthington jet changes from an upward rising jet to a descending jet, with the jet peak starting to thin, and the jet-base eventually vanishing (\textcolor{red}{S11}). When the upper sphere falls through this thinned jet, the impact with the free surface becomes more sudden, and an increase in impact acceleration occurs ($Mt \approx 4$). Eventually, the jet falls back into the pool ($Mt > 4$), the jet fully vanishes, and the upper sphere falls through the water pool almost exactly in the same manner as a quiescent drop, without any significant reduction in impact forces.

In some of the on jet cases, the Worthington jet created from pinch-off is suppressed by the splash crown dome-over at the free surface. This results in a disturbed, jet-like water column at the free surface, through which the upper sphere falls (\textcolor{red}{S10}(a)). Here, we consider these cases as `on surface jet' cases, since their formation mechanism is different from that of the on jet cases (see \textcolor{red}{SI discussion SH} and figure \textcolor{red}{S10}(a)). These cases show reduction in acceleration similar to the on jet cases as illustrated in Figure~\ref{fig8}.

\section{Conclusion}
The initial impulse force felt by any object at the initial moment of water impact can be very high (figure~\ref{fig1}c), and may prove to be catastrophic for water landing crafts or missiles, and fatal even for thrill seeking bungee jumpers \cite{vonkarman1929, mcgehee1959water, kornhauser1964,thompson1965dynamic}. Herein, we have shown through a canonical sphere impact study that the initial impact force can be greatly reduced by first launching another object in front of the body of interest. The force of impact is reduced by the cavity of the first object providing less initial water impact, lower relative velocities or upward jets that wet and decelerate the trailing body. However, if the object encounters the collapsing upward cavity in the wake of the first object, the upper sphere may experience a larger force of impact than if the leading object were not present at all.  A non-dimensional number called the Matryoshka number $Mt$ is theoretically proposed to classify two-sphere consecutive water entry behavior based on the object size and the cavity pinch-off time. Experimental results show that for $0.2< Mt < 0.75$ and $1.43<Mt<4$ significant reduction of the impact acceleration of the trailing sphere is achieved. In the interim range of $0.75<Mt<1.43$, a sudden rise in impact acceleration is seen, and must be avoided if trying to avoid catastrophic failures. This $Mt$ formulation can potentially be used to predict interactions in any multi-object water entry system, and the regime diagram proposed in figure~\ref{fig8} does well to aid in making predictions for size differences and timing.

\section{Acknowledgments}
R.R., N.S., J.B. and T.T.T acknowledge funding from the Office of Naval Research, Navy Undersea Research Program (Grant \#N000141812334), monitored by Ms. Maria Medeiros. A.K. is supported by JSPS Overseas Research Fellow Program.

\bibliography{Ref}

\begin{thebibliography}{39}
\expandafter\ifx\csname natexlab\endcsname\relax\def\natexlab#1{#1}\fi
\def\au#1{#1} \def\ed#1{#1} \def\yr#1{#1}\def\at#1{#1}\def\jt#1{\textit{#1}}
  \def\bt#1{#1}\def\bvol#1{\textbf{#1}} \def\vol#1{#1} \def\pg#1{#1}
  \def\publ#1{#1}\def\arxiv#1{#1}\def\org#1{#1}\def\st#1{\textit{#1}}

\bibitem[Aristoff \& Bush(2009)]{Aristoff2009}
{\sc \au{Aristoff, Jeffrey~M} \& \au{Bush, John W~M}} \yr{2009}  \at{{Water
  entry of small hydrophobic spheres}}.  \jt{Journal of Fluid Mechanics}
  \bvol{619},  \pg{45--34}.

\bibitem[Bergmann {\em et~al.\/}(2009)Bergmann, Van Der~Meer, Gekle, Van
  Der~Bos \& Lohse]{bergmann2009controlled}
{\sc \au{Bergmann, Raymond}, \au{Van Der~Meer, Devaraj}, \au{Gekle, Stephan},
  \au{Van Der~Bos, Arjan} \& \au{Lohse, Detlef}} \yr{2009}  \at{Controlled
  impact of a disk on a water surface: cavity dynamics}.  \jt{Journal of Fluid
  Mechanics}  \bvol{633},  \pg{381--409}.

\bibitem[Duclaux {\em et~al.\/}(2007)Duclaux, Caill{\'e}, Duez, Ybert, Bocquet
  \& Clanet]{Duclaux2007}
{\sc \au{Duclaux, V}, \au{Caill{\'e}, F}, \au{Duez, C}, \au{Ybert, C},
  \au{Bocquet, L} \& \au{Clanet, C}} \yr{2007}  \at{{Dynamics of transient
  cavities}}.  \jt{Journal of Fluid Mechanics}  \bvol{591},  \pg{177--19}.

\bibitem[Duez {\em et~al.\/}(2007)Duez, Ybert, Clanet \& Bocquet]{Duez2007}
{\sc \au{Duez, Cyril}, \au{Ybert, Christophe}, \au{Clanet, Christophe} \&
  \au{Bocquet, Lyd{\'e}ric}} \yr{2007}  \at{{Making a splash with water
  repellency}}.  \jt{Nature Physics}  \bvol{3}~(3),  \pg{180--183}.

\bibitem[Elhimer {\em et~al.\/}(2017)Elhimer, Jacques, Alaoui \&
  Gabillet]{elhimer2017influence}
{\sc \au{Elhimer, Mehdi}, \au{Jacques, Nicolas}, \au{Alaoui, A El~Malki} \&
  \au{Gabillet, C{\'e}cile}} \yr{2017}  \at{The influence of aeration and
  compressibility on slamming loads during cone water entry}.  \jt{Journal of
  Fluids and Structures}  \bvol{70},  \pg{24--46}.

\bibitem[Gekle \& Gordillo(2010)]{gekle2010}
{\sc \au{Gekle, S.} \& \au{Gordillo, J.~M.}} \yr{2010}  \at{``{G}eneration and
  breakup of worthington jets after cavity collapse. part 1. jet formation''}.
  \jt{Journal of Fluid Mechanics}  \bvol{663},  \pg{pp. 293–330}.

\bibitem[Glasheen \& McMahon(1996)]{glasheen1996vertical}
{\sc \au{Glasheen, JW} \& \au{McMahon, TA}} \yr{1996}  \at{Vertical water entry
  of disks at low froude numbers}.  \jt{Physics of Fluids}  \bvol{8}~(8),
  \pg{2078--2083}.

\bibitem[Grady(1979)]{Grady1979hydroballistics}
{\sc \au{Grady, R.~J.}} \yr{1979}  \at{Hydroballistics design handbook}.
  \jt{Naval Sea Systems command Hydromechanics Committee, January} .

\bibitem[Guillet {\em et~al.\/}(2020)Guillet, Mouchet, Belayachi, Fay, Colturi,
  Lundstam, Hosoi, Clanet \& Cohen]{guillet2020hydrodynamics}
{\sc \au{Guillet, Thibault}, \au{Mouchet, M{\'e}lanie}, \au{Belayachi,
  J{\'e}r{\'e}my}, \au{Fay, Sarah}, \au{Colturi, David}, \au{Lundstam, Per},
  \au{Hosoi, Peko}, \au{Clanet, Christophe} \& \au{Cohen, Caroline}} \yr{2020}
  The hydrodynamics of high diving.  \bt{In {\em Multidisciplinary Digital
  Publishing Institute Proceedings\/}}, ,  \vol{vol.~49},  \pg{p.~73}.

\bibitem[G{\"u}zel \& Korkmaz(2020)]{guzel2020reducing}
{\sc \au{G{\"u}zel, B{\"u}lent} \& \au{Korkmaz, Fatih~C}} \yr{2020}
  \at{Reducing water entry impact loads on marine structures by surface
  modification}.  \jt{Brodogradnja: Teorija i praksa brodogradnje i pomorske
  tehnike}  \bvol{71}~(1),  \pg{1--18}.

\bibitem[Hurd {\em et~al.\/}(2015)Hurd, Fanning, Pan, Mabey, Bodily, Hacking,
  Speirs \& Truscott]{hurd2015matryoshka}
{\sc \au{Hurd, R}, \au{Fanning, T}, \au{Pan, Z}, \au{Mabey, C}, \au{Bodily, K},
  \au{Hacking, K}, \au{Speirs, N} \& \au{Truscott, T}} \yr{2015}
  \at{Matryoshka cavity}.  \jt{Physics of Fluids}  \bvol{27}~(9),  \pg{091104}.

\bibitem[Kornhauser(1964)]{kornhauser1964}
{\sc \au{Kornhauser, M.}} \yr{1964}  \at{``{S}tructural effects of impact''}.
  \jt{Spartan Books} .

\bibitem[Korobkin \& Pukhnachov(1988)]{korobkin1988}
{\sc \au{Korobkin, A.A.} \& \au{Pukhnachov, VV.}} \yr{1988}  \at{``{I}nitial
  stage of water impact''}.  \jt{Annual Review of Fluid Mechanics}
  \bvol{20}~(1),  \pg{pp. 159--185}.

\bibitem[Li \& Sigimura(1967)]{li1967study}
{\sc \au{Li, T} \& \au{Sigimura, T}} \yr{1967}  \bt{Study of apollo water
  impact. volume 1-hydrodynamic analysis of apollo water impact final report}.
  \org{{\em Tech. Rep.\/}}.

\bibitem[Mansoor {\em et~al.\/}(2014)Mansoor, Marston, Vakarelski \&
  Thoroddsen]{mansoor2014water}
{\sc \au{Mansoor, Mohammad~M}, \au{Marston, JO}, \au{Vakarelski, Ivan~Uriev} \&
  \au{Thoroddsen, Sigurdur~T}} \yr{2014}  \at{Water entry without surface seal:
  extended cavity formation}.  \jt{Journal of Fluid Mech.}  \bvol{743},
  \pg{295--326}.

\bibitem[Marston {\em et~al.\/}(2012)Marston, Vakarelski \&
  Thoroddsen]{marston2012cavity}
{\sc \au{Marston, JO}, \au{Vakarelski, Ivan~Uriev} \& \au{Thoroddsen,
  Sigurdur~T}} \yr{2012}  \at{Cavity formation by the impact of leidenfrost
  spheres}.  \jt{Journal of Fluid Mech.}  \bvol{699},  \pg{465--488}.

\bibitem[May(1970)]{may1970review}
{\sc \au{May, Albert}} \yr{1970}  \at{Review of water-entry theory and data}.
  \jt{Journal of Hydronautics}  \bvol{4}~(4),  \pg{140--142}.

\bibitem[May(1975)]{May1975}
{\sc \au{May, A.}} \yr{1975}  \bt{``{W}ater entry and the cavity-running
  behavior of missiles''}. {\em Tech. Rep.\/}.  \org{Navsea Hydroballistics
  Advisory Committee Silver Spring Md}.

\bibitem[McGehee {\em et~al.\/}(1959)McGehee, Hathaway \&
  Vaughan~Jr]{mcgehee1959water}
{\sc \au{McGehee, John~R}, \au{Hathaway, Melvin~E} \& \au{Vaughan~Jr,
  Victor~L}} \yr{1959}  \bt{Water-landing characteristics of a reentry
  capsule}. {\em Tech. Rep.\/} No. NASA-MEMO-5-23-59L.  \org{NASA Langley
  Research Center; Hampton, VA}.

\bibitem[Moghisi \& Squire(1981)]{Moghisi1981}
{\sc \au{Moghisi, M.} \& \au{Squire, P.~T.}} \yr{1981}  \at{``{A}n experimental
  investigation of the initial force of impact on a sphere striking a liquid
  surface''}.  \jt{Journal of Fluid Mechanics}  \bvol{108},  \pg{pp. 133--146}.

\bibitem[Qi {\em et~al.\/}(2016)Qi, Feng, Xu, Zhang \& Li]{qi2016investigation}
{\sc \au{Qi, Duo}, \au{Feng, Jinfu}, \au{Xu, Baowei}, \au{Zhang, Jiaqiang} \&
  \au{Li, Yongli}} \yr{2016}  \at{Investigation of water entry impact forces on
  airborne-launched auvs}.  \jt{Engineering Applications of Computational Fluid
  Mechanics}  \bvol{10}~(1),  \pg{473--484}.

\bibitem[Seddon \& Moatamedi(2006)]{seddon2006review}
{\sc \au{Seddon, CM} \& \au{Moatamedi, M}} \yr{2006}  \at{Review of water entry
  with applications to aerospace structures}.  \jt{International Journal of
  Impact Engineering}  \bvol{32}~(7),  \pg{1045--1067}.

\bibitem[Sharker {\em et~al.\/}(2019)Sharker, Holekamp, Mansoor, Fish \&
  Truscott]{sharker2019}
{\sc \au{Sharker, Saberul~I}, \au{Holekamp, Sean}, \au{Mansoor, Mohammad~M},
  \au{Fish, Frank~E} \& \au{Truscott, Tadd~T}} \yr{2019}  \at{Water entry
  impact dynamics of diving birds}.  \jt{Bioinspiration \& Biomimetics}
  \bvol{14}~(5),  \pg{056013}.

\bibitem[Shiffman \& Spencer(1945)]{Shiffman1945}
{\sc \au{Shiffman, N.} \& \au{Spencer, D.~C.}} \yr{1945}  \bt{``{T}he force of
  impact on a sphere striking a water surface''}. {\em Tech. Rep.\/} No.
  AMG-NYU-133.  \org{Courant Institution of Mathematical Sciences, New York
  University, NY}.

\bibitem[Speirs {\em et~al.\/}(2019{\natexlab{{\em a\/}}})Speirs, Belden, Pan,
  Holekamp, Badlissi, Jones \& Truscott]{speirs2019jetball}
{\sc \au{Speirs, Nathan~B}, \au{Belden, Jesse}, \au{Pan, Zhao}, \au{Holekamp,
  Sean}, \au{Badlissi, George}, \au{Jones, Matthew} \& \au{Truscott, Tadd~T}}
  \yr{2019{\natexlab{{\em a\/}}}}  \at{The water entry of a sphere in a jet}.
  \jt{Journal of Fluid Mechanics}  \bvol{863},  \pg{956--968}.

\bibitem[Speirs {\em et~al.\/}(2019{\natexlab{{\em b\/}}})Speirs, Mansoor,
  Belden \& Truscott]{speirs_mansoor_belden_truscott_2019}
{\sc \au{Speirs, N.~B.}, \au{Mansoor, M.~M.}, \au{Belden, J.} \& \au{Truscott,
  T.~T.}} \yr{2019{\natexlab{{\em b\/}}}}  \at{``{W}ater entry of spheres with
  various contact angles''}.  \jt{Journal of Fluid Mechanics}  \bvol{862},
  \pg{R3}.

\bibitem[Speirs {\em et~al.\/}(2018)Speirs, Pan, Belden \&
  Truscott]{speirs2018water}
{\sc \au{Speirs, Nathan~B}, \au{Pan, Zhao}, \au{Belden, Jesse} \& \au{Truscott,
  Tadd~T}} \yr{2018}  \at{The water entry of multi-droplet streams and jets}.
  \jt{Journal of Fluid Mechanics}  \bvol{844},  \pg{1084--1111}.

\bibitem[Thompson(1928)]{Thompson1928}
{\sc \au{Thompson, F.~L.}} \yr{1928}  \bt{``{W}ater-pressure distribution on
  seaplane float''}. {\em Tech. Rep.\/} 290.  \org{National Advisory Committee
  for Aeronautics}.

\bibitem[Thompson(1965)]{thompson1965dynamic}
{\sc \au{Thompson, William~C}} \yr{1965} {\em Dynamic model investigation of
  the landing characteristics of a manned spacecraft\/}, ,  \vol{vol. 2497}.
  \publ{National Aeronautics and Space Administration}.

\bibitem[Truscott {\em et~al.\/}(2014)Truscott, Epps \& Belden]{Truscott2014}
{\sc \au{Truscott, Tadd~T}, \au{Epps, Brenden~P} \& \au{Belden, Jesse}}
  \yr{2014}  \at{{Water Entry of Projectiles}}.  \jt{Annual Review of Fluid
  Mechanics}  \bvol{46}~(1),  \pg{355--378}.

\bibitem[Truscott {\em et~al.\/}(2012)Truscott, Epps \&
  Techet]{truscott2012unsteady}
{\sc \au{Truscott, Tadd~T}, \au{Epps, Brenden~P} \& \au{Techet, Alexandra~H}}
  \yr{2012}  \at{Unsteady forces on spheres during free-surface water entry}.
  \jt{Journal of Fluid Mechanics}  \bvol{704},  \pg{173--210}.

\bibitem[Truscott \& Techet(2009)]{truscott2009water}
{\sc \au{Truscott, Tadd~T} \& \au{Techet, Alexandra~H}} \yr{2009}  \at{Water
  entry of spinning spheres}.  \jt{Journal of Fluid Mechanics}  \bvol{625},
  \pg{135--165}.

\bibitem[Von~Karman(1929)]{vonkarman1929}
{\sc \au{Von~Karman, T.}} \yr{1929}  \at{``{T}he impact on seaplane floats
  during landing''}.  \jt{National Advisory Committee on Aeronautics} .

\bibitem[Wang {\em et~al.\/}(2019)Wang, Faltinsen \& Lugni]{wang2019unsteady}
{\sc \au{Wang, Jingbo}, \au{Faltinsen, Odd~Magnus} \& \au{Lugni, Claudio}}
  \yr{2019}  \at{Unsteady hydrodynamic forces of solid objects vertically
  entering the water surface}.  \jt{Physics of Fluids}  \bvol{31}~(2),
  \pg{027101}.

\bibitem[Wang {\em et~al.\/}(2015)Wang, Lugni \& Faltinsen]{wang2015analysis}
{\sc \au{Wang, Jingbo}, \au{Lugni, Claudio} \& \au{Faltinsen, Odd~Magnus}}
  \yr{2015}  \at{Analysis of loads, motions and cavity dynamics during freefall
  wedges vertically entering the water surface}.  \jt{Applied Ocean Research}
  \bvol{51},  \pg{38--53}.

\bibitem[Watanabe(1933)]{watanabe1933}
{\sc \au{Watanabe, S.}} \yr{1933}  \at{``{R}esistance of impact on water
  surface, part v-sphere''}.  \jt{Scientific papers of the Institute of
  Physical and Chemical Research}  \bvol{23}~(484),  \pg{pp. 202--209}.

\bibitem[Worthington(1908)]{Worthington1908}
{\sc \au{Worthington, A~M}} \yr{1908} {A study of splashes}.

\bibitem[Worthington \& Cole(1897)]{Worthington1897}
{\sc \au{Worthington, A~M} \& \au{Cole, R~S}} \yr{1897}  \at{{Impact with a
  liquid surface, studied by the aid of instantaneous photography}}.
  \jt{Philosophical Transactions of the Royal Society of London. Series A}
  \bvol{189}~(0),  \pg{137--148}.

\bibitem[Zhao {\em et~al.\/}(2014)Zhao, Chen \& Wang]{Zhao2014}
{\sc \au{Zhao, Meng-Hua}, \au{Chen, Xiao-Peng} \& \au{Wang, Qing}} \yr{2014}
  \at{{Wetting failure of hydrophilic surfaces promoted by surface roughness}}.
   \jt{Scientific Reports}  \bvol{4}~(1),  \pg{33--5}.

\end{thebibliography}
\bibliographystyle{jfm}
\newpage
\section{Supplemental Information}

\end{document}